\documentclass[11pt]{article}  
\usepackage{graphicx}
\usepackage{amssymb,amsmath}
\newcommand{\be}{\begin{equation}}
\newcommand{\ee}{\end{equation}}
\newcommand{\bea}{\begin{eqnarray}}
\newcommand{\eea}{\end{eqnarray}}
\begin{document}
\vspace{.5in} 
\begin{center} 
{\LARGE{\bf Twisted  curve geometry  underlying topological invariants}}
\end{center} 

\vspace{.3in}
\begin{center} 
{{\bf Radha Balakrishnan$^{(1)}$, Rossen Dandoloff$^{(2)}$ and Avadh Saxena$^{(3)}$}}\\ 
{$^{(1)}$The Institute of Mathematical Sciences, Chennai 600 113, India \\
$^{(2)}$Department of Condensed Matter Physics and Microelectronics, Faculty of Physics, 
Sofia University, 5 Blvd. J. Bourchier, 1164 Sofia, Bulgaria \\ 
$^{(3)}$Theoretical Division and Center for Nonlinear Studies, Los
Alamos National Laboratory, Los Alamos, New Mexico 87545, USA}
\end{center} 

\vspace{.9in}
{\bf {Abstract:}}
  
Topological invariants such as  winding numbers and linking numbers appear 
as  charges of topological solitons 
in diverse  nonlinear physical systems described by  a unit vector field defined on two and three dimensional manifolds. While the Gauss-Bonnet theorem shows  that the Euler characteristic (a  topological invariant) can be written as the integral of the Gaussian curvature (an intrinsic geometric quantity),  the  intriguing question of whether winding and linking numbers  can also be  expressed  as integrals of some other kinds of intrinsic geometric  quantities has not been addressed in the literature. In this paper we provide the answer by  showing  that for the winding number in two dimensions,  these  geometric quantities involve  torsions of the two evolving space curves describing the manifold. On the other hand, in three dimensions we find that in addition to torsions,  intrinsic twists of the space curves  are necessary to obtain a nontrivial winding number and linking number. 
They  arise from the  hitherto unknown connections that we establish between these topological invariants and the corresponding appropriately normalized global space curve anholonomies (i.e., geometric phases) that can be associated with the unit vector fields on the respective manifolds. An application of our results to a  3D Heisenberg ferromagnetic model supporting a topological soliton is also presented.

\newpage 
  
\section{Introduction}
\label{intro}
It is by now well recognized  \cite{nakahara} that  both geometry \cite{struik}  and topology \cite{bott}  of differentiable manifolds play an important role in a variety of fields  such as condensed matter physics, high energy physics, fluid dynamics, biophysics, cosmology, etc.  
Topological invariants are quantities that take on discrete values (usually normalized to integers) that do not change under a continuous deformation of the manifold. 
 Over the past couple of decades, the study of  topological  invariants such as  winding numbers and  linking numbers  has become very valuable in the understanding of various  phenomena in diverse fields of physics, including transitions between phases with different topological properties. 

\subsection{Gauss-Bonnet relationship}
  The well known  Gauss-Bonnet theorem \cite{struik}  leads to the result that the  integral of the  Gaussian curvature $K$ over the area of a 
 compact  two-dimensional  manifold $M$  without a boundary  is a topological invariant 
 $\chi =2(1-g)$, called the  Euler characteristic, where  $g$ denotes the genus. It can be written as
 \be
\chi = \frac{1}{2\pi} \int_{M} K \,dA.
\label{GB1}
\ee
 In the above, the Gaussian curvature  $K$ is the product of the maximum and minimum curvatures on the surface at a point, given by $K = (R_{1} R_{2})^{-1}$, where $R_{1}$ and $R_{2}$ are the two corresponding  radii of curvatures. 
However, $K$  can {\em also} be expressed  entirely  in terms of the metric, i.e.,  the coefficients of the first fundamental form  of  the 2D surface, and its partial derivatives \cite{struik}. It is therefore an intrinsic   geometric quantity  inherent in the manifold,  independent of how it is embedded in space.

 The Gauss-Bonnet relationship  (1) shows  that the Euler characteristic  $\chi$ (a  topological invariant) can be written as the integral of the Gaussian curvature  $K$, an intrinsic geometric quantity. It is insightful to ask whether other topological invariants such as winding and linking numbers can also be similarly written as integrals of  certain intrinsic geometric quantities. This provides a  motivation for our paper. We will show that the answer is in the affirmative.  Interestingly, in our quest for  how to answer this question, we are led to several other new  results that give us a deeper understanding of the geometry underlying topological invariants. We  have listed these in  Sec. 3, the concluding section of this paper.


\subsection {Topological invariants (winding number and linking number) }  

To investigate whether the integrands appearing in winding and linking number expressions can be obtained in terms of some other kinds of  intrinsic geometric quantities,  we  first  scrutinize the  known integral expressions for the 2D, 3D winding and linking numbers $W_2$, $W_3$ and $H$. These are given below in Eqs. (\ref{W2}), (\ref{W3}) and (\ref{hopf-inv}), respectively. As is obvious, their integrands  typically involve angle variables, which are clearly not intrinsic geometric quantities.  Further, to understand the physical relevance of our work, it is instructive to begin with examples of physical systems where topology plays an important role.

  Topological invariants associated with a $3$-component {\em unit} vector field given by
 ${\bf t}({\bf r})$, with ${\bf r}$ representing physical space variables of the model considered, are of current interest and appear in a variety of physical systems. As examples we have the spin vector field in the continuum version of the classical  Heisenberg model with various types of interactions between  spins in magnetism \cite{kosevich,rybakov2} and the vector field  appearing in the nonlinear sigma model and its variants in high energy  physics \cite{faddeev2,gladikowski}. It is noteworthy  that  these models are generically  {\em nonlinear}, and  (under appropriate  conditions) support  {soliton  solutions for the vector field \cite{rajaraman}.  These are localized, particle-like field configurations.  (Baby) skyrmions and hopfions are well known examples of topological solitons \cite{manton,shnir} in two and three dimensions, respectively.  Topological invariants are physically interpreted as  topological charges of such solitons. 
 
  The tip of  the unit vector ${\bf t}$  lies on the two-sphere $S^2$. Consider a physical system with a {\em homogeneous}  boundary condition ${\bf t}({\bf r}) \rightarrow {\bf t}_{\infty}$,   as ${\bf r} \rightarrow \infty$,  where ${\bf t}_{\infty}$ is a constant vector.  
  Under these conditions, the  physical spaces $R^1, R^2$ and $R^3$ can be compactified to $S^1,S^2$ and $S^3$, respectively. Hence the unit vector field configurations represent maps from the (compactified) physical space to target space,  given by  $ {\bf t}(x): S^1 \rightarrow S^2$, ${\bf t}(x,y): S^2 \rightarrow S^2$, and ${\bf t}(x,y,z): S^3 \rightarrow S^2$, respectively.  In addition, we note that  using a normalized spinor representation \cite{shnir}, the target space $S^2$ can be written in terms of variables in $S^3$,  resulting in the map ${\bf T}(x,y,z): S^3 \rightarrow S^3$. 
  
  Among the above, the topological invariant of  $S^1 \rightarrow S^2$  is well known to be zero. The maps $S^2 \rightarrow S^2$ and  $S^3 \rightarrow S^3$  are classified by  integer topological invariants  called {\em winding numbers}, denoted in this paper by  $W_{2}$  and $W_3$, respectively. The former (respectively the latter) counts the number of times a sphere $S^2$ (respectively $S^3$)  wraps around another such sphere.  It is well known that the  topological solitons  associated with  these maps are baby skyrmions (respectively skyrmions).  In contrast, the map $S^3 \rightarrow S^2$ involves  manifolds with unequal dimensions. It is classified by an integer topological invariant $H$  known as the Hopf invariant.  The corresponding topological solitons are called hopfions. It can  be shown that $H$ also denotes  the {\em linking number}  of   two closed space curves in  $S^3$, which are  preimages  of any two distinct points on the target space $S^{2}$ \cite{bott}.

Using spherical polar coordinates, with polar angle $\Theta$ and azimuthal angle $\Phi$, a 3-component  unit vector field ${\bf t}$  representing $S^2$   is given by 
 \be 
{\bf t} = \big(\sin\Theta \, \cos \Phi ,\,\,  \sin\Theta  \, \sin \Phi ,\,\, 
\cos\Theta  \big ) \,, 
\label{t}
\ee
where $0 \le \Theta \le \pi$ and  $0 \le \Phi  \le  2\pi$.

The winding number $W_{2}$ in two dimensions  is given by \cite{rajaraman}   the following integral over the 2D physical manifold 
described by $y$ and $z$ (say)  
 \be
W_{2}   =
  \frac{1}{4\pi} \int\int \sin\Theta \,\left[\Theta_{y} \Phi_ z - \Theta_z \Phi_y \right]\,\, dy\,\, dz \,,  
\label{W2}
\ee
where the subscripts $y$ and $z$ on  angles $\Theta$ and $\Phi$  denote partial derivatives with respect to these variables.

 Next, we use  the following  Hopf coordinate representation \cite{shnir} to describe  a 4-component  unit vector field ${\bf T}$  representing $S^3$ 
\be
{\bf T} = (x_{1}, x_{2}, x_{3}, x_{4})
\label{T}
\ee
with
\be
x_{1}= \cos \Theta/2\, \cos {\xi_{1}};\,\, x_{2}= \cos \Theta/2\, \sin {\xi_{1}};\,\, x_{3}=\sin \Theta/2\, \cos {\xi_{2}};\,\, x_{4}=\sin \Theta/2\, \sin {\xi_{2}},
\label{hopfC}
\ee
where  $0 \le \Theta \le \pi$,  $0 \le \xi_{1} \le 2\pi$  and  $0 \le \xi_{2} \le 2\pi$.
In Eq. (\ref{hopfC}), we find it convenient to set 
\be
 \xi_{1} = (\tilde{\Psi} - \frac{\Phi}{2})\,\,; \,\, \xi_{2} = (\tilde{\Psi} +\frac{\Phi}{2}),
 \label{xi}
 \ee
giving
\be
\Phi = (\xi_{2} -\xi_{1})\,\,; \,\, \tilde{\Psi} = (\xi_{2}+\xi_{1})/2 \,. 
\label{PhiPsi}
\ee
Thus  $0 \le \Phi  \le  2\pi$, and   $\tilde{\Psi}$ represents the third angle  needed in addition to $(\Theta,\Phi)$ to define $S^3$.  Further,   $0 \le \tilde{\Psi} \le 2 \pi$.

The  winding number $W_3$  in three dimensions  is given by \cite{pak} the following integral over the 3D physical manifold 
 \be 
W_{3}  = \frac{1}{8 \pi^{2}}  \int\int\int \sin\Theta \left\{ [\Theta_y\Phi_z - \Theta_z\Phi_y] \,\, \tilde{\Psi}_x +  [\Theta_z\Phi_x - \Theta_x\Phi_z] \,\, \tilde{\Psi}_y +  [\Theta_x\Phi_y - \Theta_y\Phi_x] \,\, \tilde{\Psi}_z \right\} \, dx\,dy\, dz ,
\label{W3}
\ee 
 where the  subscripts  $x$, $y$  and $z$ on $\Theta$, $\Phi$ and 
 $\tilde{\Psi}$ stand for partial derivatives with respect to these variables.
 
 The integral expression for  the linking number $H$ (Hopf invariant)  characterizing the map $S^3 \rightarrow S^2$  is more complicated than that of $W_{2}$ and $W_{3}$.  First derived by  Whitehead \cite{whitehead},  it is  given in Eq. (\ref{hopf-inv}) below, followed by a brief discussion of how it is computed.  
 
It is important to note from Eqs.~(\ref{W2}) and  (\ref{W3}) that the integrands  of  the usual expressions for the winding numbers $W_{2}$ and $W_{3}$  are given explicitly in terms of $\Theta$ and the partial derivatives of  all the  angle coordinates of  ${\bf t}$ and  ${\bf T}$, respectively. However,  it is not at all obvious that these integrands can be expressed in terms of  geometric quantities. Needless to say,  the same holds true for  the linking number $H$ given in Eq. (\ref{hopf-inv}), whose integrand is not even given directly in terms of angle coordinates.

In the following sections, we will present an approach which shows  that the  integrands of the topological invariants $W_2$, $W_3$ and $H$ [given in Eqs. (\ref{W2}), (\ref{W3}) and (\ref{hopf-inv}), respectively] can  indeed be expressed in terms of  geometric quantities that define a space curve. We will  derive  these   quantities explicitly, and  provide their geometrical and physical interpretations.

\subsection{{\bf Methodology: Anholonomy}}  

The approach that we adopt to find the geometry underlying  winding and linking numbers involves the concept of anholonomy.  Anholonomy is a geometric phenomenon that arises if a quantity fails to recover its original value, when the parameters on which it depends are varied around a closed path in parameter space, and return to their original values. The concept of parallel transport, anholonomy and  the associated geometric phase  was first introduced by Berry in the {\em quantum}  context \cite{berry}. However, as pointed out by Berry in \cite{shapere},  there is a direct correspondence between the anholonomy of a {\em classical vector} and that  of a quantum state. Hence in the present paper which deals with {\em classical}  anholonomy,  we will use the quantum terminology such as  Berry connection, Berry phase, Berry curvature, etc., for their classical analogs as well. 

 Our methodology is summarized as follows: We map the 3-component unit vector field ${\bf t}$ to the tangent of a space curve. 
 We define a  generalized Frenet-Serret frame for the space curve \cite{struik}, where the plane composed of the normal and binormal vectors of the frame  is {\em rotated} (i.e., {\em  twisted}) around the tangent to the curve, by an arbitrary angle. We invoke the concept of  parallel transport and find the associated anholonomy density (i.e., Berry connection) \cite{berry,shapere} of  the space curve depicting a 1D manifold. We then describe  the 2D and 3D manifolds in terms of appropriate evolving space curves and find the associated global anholonomies. This leads to
  new relationships between winding and linking numbers and the corresponding  (appropriately normalized) global anholonomies associated with the unit vector fields on the respective manifolds.
 These hitherto unknown connections help us to demonstrate that these topological invariants  can be  expressed as integrals involving  certain  geometric  quantities  appearing in the intrinsic description of a space curve. These will be  found explicitly and interpreted physically in the next section. 
  
\section {Determination of topological  invariants as integrals involving  geometric quantities of a space curve}
  To set the stage for higher dimensions, it is instructive to begin with  a  one-dimensional  manifold. 
  
\subsection {One-dimensional manifold}
 The following  important observation is in order.  As explained in Sec. 1.2, our starting point is a (given)   3-component {\em unit}  vector ${\bf t}$. Let us  associate ${\bf t}$ with the tangent of  a three-dimensional  space curve  ${\bf R}(x)$ parametrized by $x$, such that ${\bf t}(x) = d{\bf R}(x)/dx $. Since  ${\bf t}$ is a unit vector, this in turn implies that the  {\em arc length} of the space curve, which is  defined  \cite{struik}  by  $s = \int |d{\bf R}(x)/dx | \,dx$  yields  $x$.  In other words, the arc length variable $s$  of the space curve gets  directly identified with the physical coordinate variable $x$.  
    
  The usual   Frenet-Serret equations for  the orthogonal unit vector triad $({\bf t}, {\bf n}, {\bf b})$ defined on a space curve with arc length $x$ are given by \cite{struik}
 \be
{\bf  t}_{x} = k_1  {\bf n}\, ; ~~
{\bf n}_{x} = -k_1 {\bf t} + \tau_{1} {\bf b} \,; ~~
{\bf b}_{x} = -  \tau_{1}  {\bf n} \,.
\label{FS1}
\ee
 In Eq. (\ref{FS1}), the subscript $x$ on the various vectors  denotes derivative with respect to $x$.
 The principal normal vector ${\bf n}$ is {\it defined} to be along 
${\bf t}_x$.  The binormal vector is given by ${\bf b} ={\bf t}\times {\bf n}$.  Further,  $k_1$  is the {\em curvature}, which measures the departure of a curve from a straight line, while 
 $\tau_{1}$ is  the {\em torsion}. It is a measure of the nonplanarity of  the curve, i.e., the degree to which the curve  twists out of a plane. 
 As is well known \cite{struik}, curvature $k =k_{1}(x)$ and torsion $\tau =\tau_{1}(x)$  are the {\em intrinsic} equations of the space curve.
 \begin{figure}[h] 
\centering 
\includegraphics[width=2.3 in]{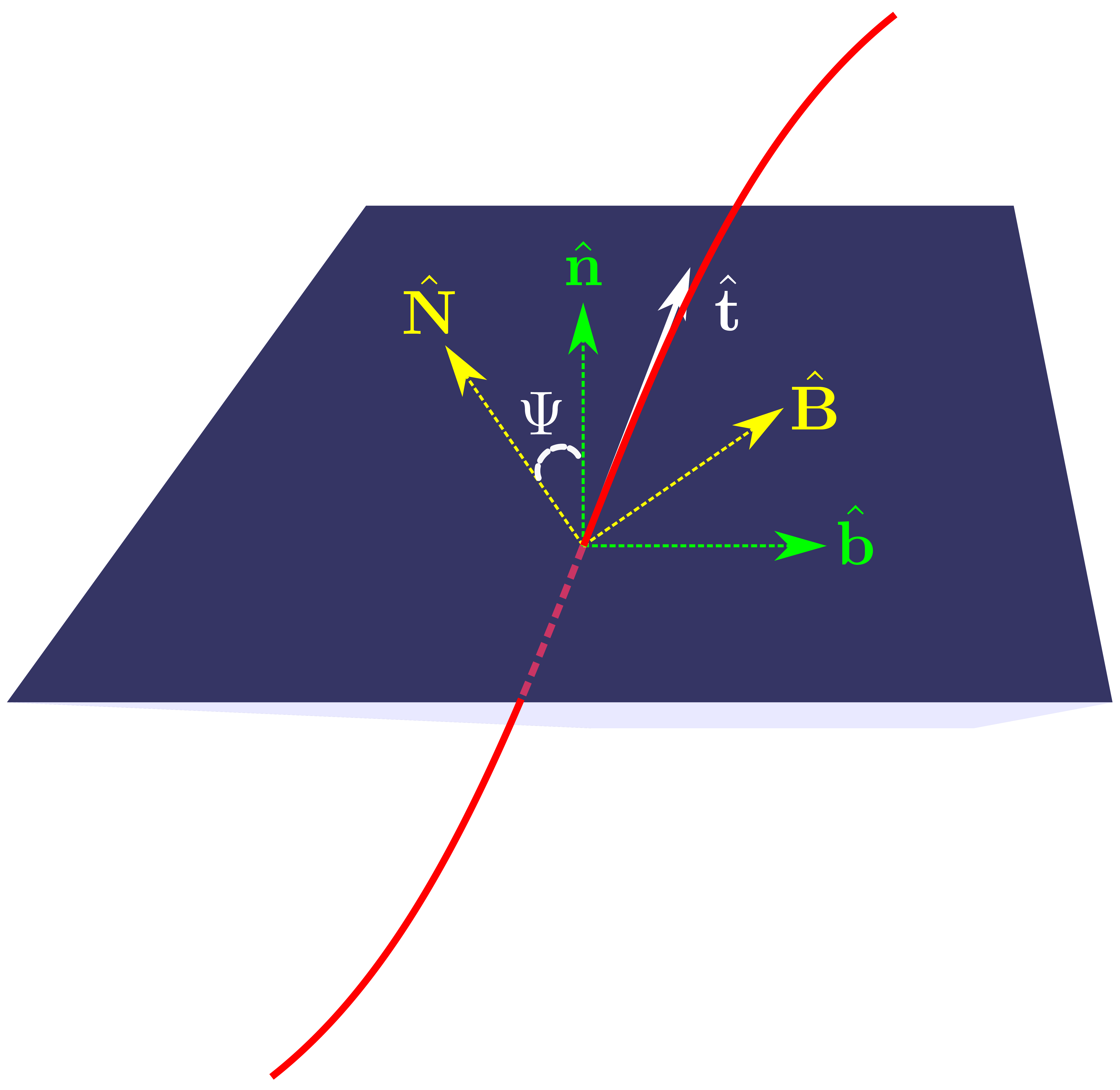} 
\caption{A  space curve characterized by the unit triad $(\hat{\bf t}, \hat{\bf n}, \hat{\bf b})$ representing the tangent, normal and binormal to the curve.  More generally, the curve can be represented by a rotated triad $(\hat{\bf t}, \hat{\bf N}, \hat{\bf B})$, with the angle of rotation of the $(\hat{\bf n}, \hat{\bf b})$ plane around 
$\hat{\bf t}$ given by $\Psi$.}
\end{figure} 

Clearly, one has the freedom to  choose new unit vectors $({\bf N}, {\bf B})$  perpendicular to ${\bf t}$ (Fig. 1),  
  by rotating \cite{moffatt2}  the Frenet pair $({\bf n},{\bf b})$ around the tangent  by an angle $\Psi$, with $0 \le \Psi \le 2 \pi$.  
This leads to  
 \be
 {\bf N} = {\bf n} \cos \Psi +{\bf b} \sin \Psi;\,\,\, ~{\bf B}  = {\bf t} \times {\bf N} = -{\bf  n} \sin \Psi +{\bf b} \cos \Psi .
 \label{NB}
 \ee
 Here, $\Psi$ can be physically interpreted as  the local {\it intrinsic twist}  of the $({\bf N}, {\bf B})$ plane around the  (tangential) axis of the space curve. This inherent freedom in the choice of  $\Psi$  can also be regarded as a  gauge freedom. As we will demonstrate,  $\Psi$  plays a crucial role in unraveling the underlying geometry of  certain topological invariants. 

After a straightforward calculation using Eq. (\ref{NB}), the usual Frenet-Serret equations given in Eq. (\ref{FS1})
can be written in terms of  the {\em rotated}  orthogonal unit triad $({\bf t}, {\bf N}, {\bf B})$  (see Fig. 1) as: 
\be
{\bf  t}_{x} = (k_1 \cos \Psi) {\bf N} - (k_1 \sin \Psi) {\bf B} \,, ~~~
{\bf N}_{x} = -(k_1 \cos \Psi) {\bf t} + \tau_{T,1} {\bf B} \,, ~~~
{\bf B}_{x} = (k_1 \sin \Psi ) {\bf t}  -  \tau_{T,1}  {\bf N} \,. 
\label {FSt}
\ee 
(For notational convenience,  the subscripts $1$, $2$ and $3$  for the curve parameters $k$, $\tau$ and $\tau_{T}$ will be  used to indicate  that they correspond  to a  space curve parametrized by $x$, $y$ and $z$, respectively.) 

In Eq. (\ref{FSt}),  the curvature $k_{1} =|\mathbf{t}_{x}|$, and  $\tau _{T,1}$ is given by 
\be
\tau _{T,1} = \tau _ {1} + \Psi_{x} \, ,
\label{twist}
\ee
where  $\tau_{1}$ is the  {\it torsion}  that appears in the usual {\em unrotated}  Frenet-Serret equations ({\ref{FS1}) and  $ \Psi_x$ denotes  derivative of $\Psi$ with respect to $x$.
As is well known \cite{struik}, $\tau_{1}$  is given by 
\be 
\tau_{1} =\mathbf{t}~\cdot~
(\mathbf{t}_{x}~\times~\mathbf{t}_{xx})~/~k_{1}^{2}.
\label{tau1}
\ee
Adopting the nomenclature used  by Moffatt and Ricca \cite {moffatt2}  in  fluid dynamics,
 we  refer to  $\tau_{T,1}$   given in  Eq. (\ref{twist}) as the {\it total twist density} of  a space curve with arc length $x$. This is because 
$(1/2\pi) \int \tau_{T,1}  dx$   has 
been called  the  `total twist number'  ($Tw$)  of a  thin ribbon constructed from the space curve using a standard procedure \cite{calug,fuller}.
We  call the term $\Psi_{x}$ appearing  in Eq. (\ref{twist}) as  the  {\em intrinsic  twist density},  since its integral
\be
(1/2\pi) \int \,\,\Psi_{x} \,\, dx =\, \mathcal{N}
\label{scriptN}
\ee
is called the {\em twist number}.
On integrating Eq. (\ref{twist}), we find  that the total twist number is the sum of the total torsion and the twist number,  as was defined in \cite{moffatt2}.

Returning to Eqs. (\ref {FSt}), we note that the three equations for the triad $({\bf t}, {\bf N}, {\bf B})$ can be combined to give  ${\bf G}_{x} =  \boldsymbol {\xi}  \times {\bf G}$,
 where ${\bf G}$ stands for ${\bf t}, {\bf N}$  or ${\bf B}$. 
 In the above, the  Darboux vector \cite{struik} $\boldsymbol {\xi}$  that denotes the angular velocity of the  rotation of the triad  as one moves along the curve is given by
\be 
\boldsymbol {\xi} =  \tau_{T,1}\,\,  {\bf t}  + (k_{1} \sin \Psi )\,\, {\bf N} +\,( k_{1} \cos \Psi) \,\,{\bf B} \,. 
\ee
A non-rotating plane can be defined by using the usual  parallel transport  \cite{kugler, BBD1} of the tangent vector ${\bf t}$ along the space curve in such a way that as $x$ changes to $(x + dx)$, the $({\bf N}, {\bf B})$ plane gets rotated by an infinitesimal angle  $\tau_{T,1} \, dx $  with respect to a non-rotating plane. This  signifies an associated {\em anholonomy density} \cite{berry}  (with angle per unit length as dimension) for the 1D manifold we are considering. It is the {\em classical}  analog of  {\em Berry connection}, which we denote by $V_{1}$. It is given by
\be
 V_{1} = \tau_{T,1}. 
 \label{V1}
 \ee
 Eq. (\ref{twist}) shows that   $V_{1}$  is a  gauge dependent quantity,  since  the rotation angle $\Psi(x)$ can be arbitrary.
 
 It is instructive to express $\tau_{1}$  in terms of spherical polar coordinates of ${\bf t}$,
 by substituting  Eq. ({\ref{t}) in Eq. (\ref{tau1}).  Using $k_{1} =|\mathbf{t}_{x}| =\Theta^{2}_{x} + \sin^{2}\Theta \,\, \Phi^{2}_{x}$ in the latter equation, a short calculation yields 
  
\be
 \tau_{1} =\mathbf{t}~\cdot~
(\mathbf{t}_{x}~\times~\mathbf{t}_{xx})~/~k_{1}^{2} = \cos \Theta \,\, \Phi_{x} +   \beta _{x},
\label{tau1angles}
\ee
where
\be
\beta =  \tan^{-1}  \left(\frac {\sin\Theta\,\, \Phi_{x}}{\Theta_x}\right).
\label{beta}
 \ee
 The subscripts $x$  on the various quantities denote their $x$-derivatives. 

Using  Eq. (\ref{tau1angles}) in  Eq. (\ref{twist}) and substituting it in  Eq. ({\ref{V1}), we obtain  the space curve anholonomy density for the 1D manifold (described by the space curve with arc length $x$), i.e., the Berry connection,  to be
\be
 V_{1} = \tau _{T,1}= \cos \Theta\,\, \Phi_{x}\, + \,\,\tilde {\Psi}_{x} \,,
 \label{twist_Psi}
 \ee
 where we have defined  $\tilde{\Psi}_{x} = \Psi_{x} + \beta_{x}$.
 
 It is important to note that $\beta$ appears as a {\em total derivative}  in  $V_1$.
  If the final direction of  the tangent of the space curve is the same as its initial direction,  the spherical image of  ${\bf t}(x)$ traces a {\em closed path} denoted by $C$ on $S^2$. When $V_1$ is integrated over $C$, it is clear that in this case, the  contribution from $\beta$  {\em vanishes}. With $\Psi$ also satisfying similar  cyclic boundary conditions, we find that the integral of ${\tilde{\Psi}_{x}}$  yields  
 [${\Psi}_{final}-{\Psi}_{initial}] =2 \pi \mathcal{N}$, where $\mathcal{N}$ is an integer counting the number of twists the  curve (thin ribbon) undergoes while traversing the  closed path $C$.}  Hence, on substituting Eq. (\ref{twist_Psi}) in  Eq. (\ref{V1})  and integrating it, we obtain the  {\em total 1D global  anholonomy}    $Q_{1}$ or {\em Berry's geometric phase} to be
 \be
 Q_{1} =  \oint _{C} \, V_{1}\, dx = \oint_{C} \tau _{T,1} \, dx = -\oint_{S} \, \sin \Theta\, d\Theta\, d\Phi + 2 \pi \mathcal{N}\, = \Omega(C) + 2\pi \mathcal{N},
 \label{Q1}
 \ee
 where $\Omega(C)$  is the solid angle subtended by the closed boundary curve $C$ (enclosing a surface $S$) at the center of the sphere.  Thus the Berry phase  $Q_{1}$ is independent of the function $\Psi(x)$ representing the rotation given in Eq. (\ref{NB}). 
 
 \subsection{Two-dimensional manifold}
 
 Next, we consider a 2D manifold parametrized by $(y,z)$,  
  with a vector field ${\bf t} (y,z)$ defined on it.
  Here, for  every {\it constant}  $z$ (resp. $y$) on the 2D manifold, there is a space curve parametrized by $y$  (resp.$z$), with its unit tangent  denoted
 by ${\bf t}(y)$ (resp. ${\bf t}(z)$). Since ${\bf t}$ is a unit vector, the physical coordinate $y$ (resp. $z$) can be regarded as the arc length variable, as explained above Eq. (\ref{FS1}).  We can write down two sets of Frenet-Serret equations  in terms of a {\em rotated} unit triad (analogous to  Eq. (\ref{FSt})) to describe this 2D case.  As mentioned  below Eq. (\ref{FSt}),  in each set, we  now have   $k_{1}$ replaced by $k_{2}$ (resp. $k_{3})$ and  $\tau_{T,1}$ replaced by
 $\tau_{T,2}$  (resp. $\tau_{T,3}$), and the subscript $x$  replaced by $y$ (resp. $z$).
 
 It has been shown in \cite{BBD1} that  when one considers  a  closed path of  infinitesimal area $dy \,dz$ on the two-dimensional manifold  parametrized by $(y,z)$, and  invokes the concept of  parallel transport  of ${\bf t}$  (as in the 1D case), then, on returning to its starting point on the curve, the ({\bf N}, {\bf B}) plane gets rotated  by an angle
 $[\partial \tau_{T,3}/\partial {y}  - \partial \tau_{T,2}/\partial {z}]\,\, dy\, dz$, which is  just a measure of the associated anholonomy. This derivation is similar to that given in  \cite{BBD1} (except that  instead of the usual  {\em unrotated}  Frenet-Serret equations used there, we now use Eqs. (\ref{FSt})), and  will not be repeated here.
 
 We denote the 2D {\it anholonomy density}  (with angle per unit area as dimension) associated with the  2D manifold ($y,z$)  by  $\Omega_{1}(y,z)$. It is given by 
\be 
\Omega_{1}(y,z)=  \left[\frac{\partial \tau_{T,3}}{\partial{y}}  - \frac {\partial  \tau_{T,2}}{\partial{z}}\right] \,. 
\label{O1}
\ee
Equation ({\ref{O1}) is reminiscent of  a {\em Berry curvature}, since it has  the form of the  $x$ component of  the curl of a  certain `vector potential'  whose components are torsions of the two curves, i.e.  the $x$ component of a  (fictitious)  `magnetic field'.

On substituting  (the analogs of) Eq. (\ref{twist}) in Eq. (\ref{O1}), we find that the terms involving $\Psi$ cancel out. Thus we obtain the {\em 2D anholonomy density}  to be 
\be
\Omega_{1}(y,z) =  \left[\frac{\partial \tau_{3}}{\partial{y}}  - \frac {\partial  \tau_{2}}{\partial{z}}\right] .
\label{O1tau}
\ee
Note that $\Omega_{1}(y,z)$  becomes {\em independent of $\Psi$}, and is hence {\em gauge invariant}.

   The  total anholonomy  $Q_{2}$ associated with the vector field ${\bf t}(y,z)$ is given by integrating  the above  expression for the anholonomy density over the 2D manifold, yielding 
 \be
Q_{2}  =  \int\int \Omega_{1}(y,z)  dy dz = \int\int \left[\frac{\partial \tau_{3}}{\partial{y}}  - \frac {\partial  \tau_{2}}{\partial{z}}\right] dy\, dz\,.
\label{Q2}
\ee

On using (the analogs of) Eq. (\ref{tau1angles})  to write down   $\tau_{2}= \cos \Theta \,\,\Phi_{y} +\beta_{y}$ and 
$\tau_{3}= \cos \Theta \,\,\Phi_{z} +\beta_{z}$ in Eq. (\ref{O1tau}),  we find that the terms involving 
 $\beta$ cancel, and  the expression for the  2D 
anholonomy density or Berry curvature becomes
\be
\Omega_{1}(y,z) = \left[\frac{\partial \tau_{3}}{\partial{y}}  - \frac {\partial  \tau_{2}}{\partial{z}}\right] =
 \sin\Theta \,\left[\Theta_{y} \Phi_ z - \Theta_z \Phi_y  \right] ,
\label{O1angles}
\ee
where the subscripts on the angles $\Theta$ and $\Phi$ denote partial derivatives.
Substituting Eq. (\ref{O1angles}) in Eq. (\ref{Q2}), we find the total anholonomy $Q_2$ (which is just  the integral of the Berry curvature) to be
\be
Q_{2} =  \int\int \sin\Theta \,\left[\Theta_{y} \Phi_ z - \Theta_z \Phi_y \right]\, dy dz \,.  
\label{Q2W2}
\ee
 
 Now, it  is important to note that Eq. (\ref{Q2W2}) is independent of $\Psi$, and also remains valid when we consider a non-rotated frame with $\Psi =0$. Thus it becomes possible to relate $Q_2$ to the winding number $W_2$ given in Eq. (\ref{W2}). We obtain
 \be
 W_{2}= \frac{1}{4\pi} \,\, Q_{2}.
 \label{WQ2}
 \ee
 This  yields a direct  relationship between the winding number $W_2$ and  the global 2D anholonomy $Q_2$.
   
 
 Further, substituting Eq. (\ref{Q2}) in Eq. (\ref{WQ2}) , we obtain
 \be
W_{2}  =  \frac{1}{4\pi} \int\int \left[\frac{\partial \tau_{3}}{\partial y}  - \frac {\partial \tau_{2}}{\partial z}\right] dy dz.
 \label{W2torsion}
\ee
 {\em Our new results that determine the  geometric quantities underlying the 2D winding number $W_{2}$ are summarized as follows:}
 
 \noindent {\bf (i)} Equation (\ref{WQ2}) shows the following new connection:  The winding number $W_2$,  which is the  topological invariant of the map $S^2 \rightarrow S^2$,  is {\em identical} to the  global anholonomy $Q_{2}$ associated with  evolving space curves depicting a  2D manifold (which is also the integral of the Berry curvature)  with a normalization factor ($1/4\pi$). \\
 \noindent {\bf (ii)} Equation (\ref{W2torsion}) shows that the {\em integrand} of the topological invariant $W_{2}$  is  expressed in terms of the {\em torsions}  of the two space curves forming the 2D physical manifold. These torsions represent  the geometric quantities we were looking for in the integrand of the topological invariant $W_2$. Note that they  describe  {\em intrinsic  curve geometry}, and are physically interpreted as the {\em nonplanarity} of the space curves involved.

\subsection{Three dimensional manifold}

 We may regard the 3D manifold  to be generated by a  2D manifold parametrized by $(y,z)$  with a vector field 
 ${\bf t} (y,z)$ defined on it  (see Sec. 2.2), which evolves as the third parameter $x$ varies. We can equivalently {\em also} have the  two other possibilities, namely, the $(z,x)$  manifold evolving with $y$, and the $(x,y)$ manifold evolving with $z$. As explained in Secs. (2.1) and (2.2), since ${\bf t}$ is a {\em unit}  vector,  the physical coordinates $x,y$ and $z$ of the manifold get identified with arc length variables for the corresponding curves. 
 
 \subsubsection{Winding number $W_3$ }
 
 It is convenient to start with the integral expression of $W_3$ given in Eq. (\ref{W3}) 
  and write it in the form
 \be
W_3= \frac{1}{8\pi^2}\,\, Q_3 \,,
\label{W3Q3}
\ee
where
\be
Q_3 = \int\int\int \sin\Theta \left\{ [\Theta_y\Phi_z - \Theta_z\Phi_y] \,\, \tilde{\Psi}_x +  [\Theta_z\Phi_x - \Theta_x\Phi_z] \,\, \tilde{\Psi}_y +  [\Theta_x\Phi_y - \Theta_y\Phi_x] \,\, \tilde{\Psi}_z \right\} \, dx\,dy\, dz \,.
\label{Q3angles}
\ee
Note that the first  term  in the integrand of $Q_3$  is given by $ \sin\Theta \,\, (\Theta_y\Phi_z - \Theta_z\Phi_y)] \,\, \tilde{\Psi}_x$.  We immediately realize  [from Eqs.  (\ref{O1angles}) and (\ref {twist_Psi})] that this term
 can be written as  a {\em product}  of 2D anholonomy density/Berry curvature $\Omega_{1} (y,z)$  associated with the $(y,z)$ manifold,   and the corresponding 1D anholonomy density/Berry connection given by $V_1$ for the  1D manifold  (created by the $x$ space curve),  {\em provided}  the term  $\sin \Theta  (\Theta_y \Phi_z -\Theta _z \Phi_y) \cos \Theta \,\Phi_{x}$ is {\em subtracted} from it.
The other two terms in the integrand of $Q_3$ can also  be similarly written down, with appropriate terms  subtracted in an analogous fashion. 
A short calculation shows that the total contribution from the three subtracted  terms  {\em  vanishes}  exactly.

Hence the expression (\ref{Q3angles})  for $Q_3$ can be written in a compact form as
\be 
Q_{3} =   \int\int\int {\bf V}. \, \boldsymbol{\Omega} \,\, dx\,\, dy\,\, dz\,\, ,
\label{Q3Omega}
\ee
where  the  vectors ${\bf V}$ and 
$\boldsymbol{\Omega}$  are defined as 
\be 
{\bf V} = (V_1, V_2, V_3) =  (\tau_{T,1}, \tau_{T,2}, \tau_{T,3})= \boldsymbol {\tau}_{T} \,,
\label{Vtau}
\ee 
and
\be 
\boldsymbol{ \Omega} =  \boldsymbol{\nabla}\times{\bf V} = \boldsymbol{\nabla}\times\boldsymbol{\tau}_{T} \,.
\label{OmCross}
\ee 
The  components  $V_{i}$ and $\Omega_{i}$, for  $i=2,3$  are obtained analogous to  $V_1$ and $\Omega_1$ defined in Eq. (\ref {V1}) and Eq. (\ref{O1}), respectively. 
Further, Eq. (\ref{OmCross}) yields  $\boldsymbol{\nabla}\cdot \boldsymbol{ \Omega} = 0$, showing that $\boldsymbol{ \Omega}$ is a divergence-free vector field, reminiscent of a magnetic field.  

 It is evident that  the 3D integral for $Q_3$ given in Eq.~(\ref{Q3Omega})  encodes the  global anholonomy  of the 3D manifold, since its integrand is given in terms of the respective Berry connections $V_{i}$ and  the Berry curvatures $\Omega _{i}$, $ i=1,2,3$,  that emerge in the three possible descriptions of the 3D manifold.
 
 To determine the geometric quantities in the integrand of $W_3$, we first use 
 Eqs. ({\ref{Vtau}) and (\ref{OmCross}) in Eq. ({\ref{Q3Omega}), and write the 3D global anholonomy as
\be 
Q_{3} =  \int\int\int\, \boldsymbol{ \tau}_{T} \cdot (\boldsymbol{\nabla} \times \boldsymbol {\tau}_{T} ) \,\, dx\,\, dy\,\, dz \,.
\label{Q3tauT}
\ee
In Eq. (\ref{Q3tauT}),  the $x,y$ and $z$ components of $\boldsymbol{ \tau}_{T}$ are given by Eq. (\ref{twist}) and its  analogs.  

Substituting  Eq. (\ref{Q3tauT}) in Eq. (\ref{W3Q3}) yields 
\be
W_{3} = \frac{1}{8\pi^2}  \int\int\int\, \boldsymbol{ \tau}_{T} \cdot (\boldsymbol{\nabla} \times \boldsymbol {\tau}_{T} ) \, dx\, dy\, dz .
\label{W3tauT}
\ee
\noindent {\em We  summarize  below, our new  results that determine the  geometric quantities  for  the 3D winding number $W_3$:} \\
\noindent {\bf (i)} Equation (\ref{W3Q3}) shows that the  3D winding number $W_3$ given above,  which is the  topological invariant of the map $S^3 \rightarrow S^3$,  is  {\em identical} to  the   global  anholonomy  $Q_3$  (see Eq. (\ref{Q3tauT})) associated with the evolving space curves  in a 3D manifold, with a normalization factor $(1/8\pi^2)$.  

 \noindent {\bf (ii)} Note that the three components of  $\boldsymbol{ \tau}_{T}$ appear in the winding number $W_3$ given in Eq. (\ref{W3tauT}). They are given
 analogous to the first component  $\tau _{T,1} = \tau _ {1} + \Psi_{x}$ defined  in Eq. (\ref{twist}).  Thus, we see that the topological invariant $W_3$ can be written as an integral of  terms involving the torsions (a measure of nonplanarity)  of the evolving space curves depicting the  3D manifold, as well as  the  respective  derivatives of the  twist $\Psi$.  Both these are intrinsic
  geometric quantities of a space curve.  In particular, we note that   in contrast to the 2D winding number $W_2$,  the intrinsic twist $\Psi$  of the space curve that appears in  $\tau _{T,1}$   is {\em necessary}  to yield a nontrivial winding number $W_3$. 
  
 Although we have not used the spinor representation of  the unit vector field ${\bf t}$ in this paper,
 we present a short discussion on it, as a brief digression. As seen from  Eq. (\ref{NB}), the twist $\Psi$  encodes a certain {\em gauge freedom}   in the choice of the $({\bf n}, {\bf b})$ plane of the Frenet-Serret frame of a space curve. We wish to  determine the  role of the  twist 
 $\Psi$, in the following  normalized spinor representation \cite{shnir} for  the unit vector ${\bf t}$  given in  Eq. (\ref{t}):
\be
{\bf t} = Z^{\dagger}\, \boldsymbol{\sigma}\, Z \,.
\label{tSigma}
\ee
Here, the three  components of  $\boldsymbol{\sigma}$ are the well known Pauli spin matrices. Setting $Z^{T} = (z_{1}, z_{2})$, with $z_{1} = x_1+i x_2$ and $z_2 = x_{3}+ ix_{4}$, the condition  $(|z_{1}|^2 + z_{2}|^2)=1$ implies that the real coordinates $x_{i}$, $i=1,2,3,4$,  lie on $S^3$. 

Next, we use the Hopf coordinate representation for $S^3$ given in Eq. (\ref{hopfC})
to give
\be
z_{1} = \cos \Theta/2\,\, \exp{i\xi_{1}} \,\,;  \,\,\,z_{2} = \sin \Theta/2\,\, \exp{i\xi_{2}} \,. 
\label{z1z2}
\ee
Substituting  the definitions from Eq. (\ref{xi}) in $Z^{T} = (z_{1}, z_{2})$, we get
\be
Z^{T} = \exp (-i{\tilde{\Psi}})\,  \left( \cos {\Theta/2}\, \exp({-i\Phi/2}) \,, \, \sin{ \Theta/2}\,\, \exp({i\Phi/2}) \right ). 
\label{gauge}
\ee
 Integration of the derivative relationship given below Eq. (\ref{twist_Psi}) shows that the angle 
 $\tilde{\Psi}  = \Psi + \beta +C_0$, where $C_0$ is an arbitrary  constant. Thus the intrinsic twist $\Psi$, which can be regarded as a gauge freedom inherent  in the choice of the rotation angle $\Psi$ in the space curve formulation, appears in  the gauge freedom  $\tilde{\Psi}$  inherent in the spinor representation  $Z$ of ${\bf t}$  given in  Eq. (\ref{gauge}).}

 \subsubsection{Linking number (Hopf invariant $H$)}
Since the Hopf invariant $H$ characterizes the map $S^3 \rightarrow S^2$ between two manifolds of unequal dimensions,
 it  is not a winding number.  It can  be shown to be the {\em linking number}  of the  two closed space curves in $S^3$,  that are the  preimages  of any two distinct points on the target space $S^{2}$.
The integral expression for  $H$  was first derived by Whitehead \cite{whitehead}, and can be written in the form \cite {gladikowski}
\be 
 H = \frac{1}{8\pi^{2}} \iiint (\mathbf{A} \cdot \mathbf{B}) \,
 dx\,  dy\, dz,
 \label{hopf-inv}
\ee
where the Cartesian components $(B_1,B_2,B_3)$ of the divergence-free field  ${\bf B}$  (called {\em emergent magnetic field} in magnetic models)   
are given by
\be
B_{1}  = \mathbf{t} \cdot(\partial_{y}\mathbf{t}\times\partial_{z}\mathbf{t}) = \sin\Theta \,\left[\Theta_{y} \Phi_ z - \Theta_z \Phi_y  \right],
\label{B1}
\ee
with the components $B_{2}$ and $B_{3}$ given in terms of its cyclic permutations. (The last equality in Eq. (\ref{B1}) is obtained by using Eq. (\ref{t}) for ${\bf t}$.) It is easily  verified  that $\nabla\cdot \mathbf{B} =0$.

Now, by solving for ${\bf t}({\bf r})$  in any physical model in 3D, the components of ${\bf B}$ can be explicitly written down using Eq. (\ref{B1}) and its analogs.  Given the vector  ${\bf B}$, its vector potential  $\mathbf{A}$  is found by solving  
\be
\boldsymbol{\nabla}\times\boldsymbol{\bf  A} ={\bf B} \,.
\label{solveA}
\ee
Next, the solution for ${\bf A}$ and the corresponding ${\bf B}$  are  substituted in Eq. (\ref{hopf-inv}), and the integral computed,  to determine the Hopf invariant $H$.

In order to identify the curve geometry underlying the topological invariant $H$, we begin by observing  that the integral in Eq. (\ref{hopf-inv}) has the {\em same form} as the anholonomy $Q_3$ given in Eq. (\ref{Q3Omega}), with ${\bf A}$  and ${\bf B}$ identified with ${\bf V}$ and  $\boldsymbol\Omega$, respectively.

  We note from Eq. (\ref{B1}) that ${\bf B}$ is given  in terms of  spherical coordinates $(\Theta,\Phi)$ of ${\bf t}$, and their partial derivatives. Therefore the  vector  potential  ${\bf A}$, which is  found by solving  Eq. (\ref{solveA}) will also  depend  only on $(\Theta,\Phi)$. 
  On the other hand, the space curve anholonomy density  (or Berry connection)  ${\bf V}$ given in Eq. (\ref{twist_Psi}) shows that the solution for the components of the vector potential must satisfy
 \be
 A_{i}(\Theta, \Phi)=V_{i} = \tau_{T,i} = \tau_{i} + \partial_{i} \Psi,
 \label{AThetaPhi}
 \ee  
  where for $i=1,2,3$,  the term $\partial_{i} \Psi$  is given by  the partial derivatives $\Psi_{x}$,  $\Psi_{y}$ and $\Psi_z$, respectively. Since the components $\tau_{i}$ depend on  (appropriate partial derivatives of) 
   $(\Theta,\Phi)$ [see Eq. (\ref{tau1angles})],  Eq. (\ref{AThetaPhi})  shows that the twist 
 $\Psi$  will also become a functional  of  these two angles. 
 
 Hence $\Psi$ is not an independent angle in the expression for the general 3D anholonomy given in Eq.~(\ref{W3Q3}), but gets {\em fixed} in terms of  some functions of $\Theta$ and $\Phi$, in general. (This is similar to `gauge-fixing'  or choosing a gauge, as encountered in other physical contexts such as electromagnetism.) 
 
 Due to the above gauge fixing,  the relationship  given in Eq. (\ref{W3Q3}) can be written as
 \be
H = (1/8\pi^2)\,\, Q_{3,F},
\label{HQ3F}
\ee
where the integral expression for $Q_{3,F}$  has the {\em same form}  as  Eq. (\ref{Q3tauT}), 
 but with  $\Psi$ appearing in its integrand getting {\em fixed} as a {\em function of}  the spherical angles of ${\bf t}$ representing $S^2$, as already explained. (Hence in Eq. (\ref{HQ3F}), we have used the subscript $F$ on $Q_3$  to denote `fixed $\Psi$'.)

It is noteworthy that the presence of  intrinsic twists $\Psi$ of the space curves, although fixed in this case,  is still {\em necessary}  to obtain a nontrivial Hopf invariant, just as it was in the case of the winding number $W_3$. 
 Clearly, the functional form of  the twist $\Psi$ will depend on the vector field solution ${\bf t}(\Theta, \Phi)$ supported by  the physical model considered. As an example, we present  a 3D ferromagnetic model.
 \vskip 0.3truecm 
 \subsection {An application to a magnetic model: Determination of intrinsic twists in the Hopf invariant} 
 
 It should be  clear from Eq. (\ref{AThetaPhi}) that to determine an  analytical expression for  the intrinsic twist $\Psi$ appearing in the integral expression of the Hopf invariant $H$ for any physical  model  in three dimensions,  we require analytical solutions  for its unit vector field  configurations
 ${\bf t}(\Theta,\Phi)$.  Here, we point out that several  3D models  have  been considered in the literature over the past two decades,  with
   homogeneous boundary conditions (see Sec.~1.2) imposed on the solutions for ${\bf t}(x,y,z) $.  These describe the map $S^3 \rightarrow S^2$, as appropriate for  finding  a {\em hopfion}  configuration with an integer Hopf invariant.  However, to our knowledge, only 
   {\em numerical solutions}  for ${\bf t}$  (as well as $H$) have been found so far (see, for e.g.,  \cite{ rybakov2, faddeev2,gladikowski,manton, shnir}). This is essentially because
the Euler-Lagrange variational equations  for ${\bf t}$  derived from the  energy  expressions (which as is well known, are coupled nonlinear partial differential equations for $\Theta$ and $\Phi$)  have been found to be  difficult to solve analytically  in these models, rendering them unsuitable for finding  an  analytical expression for $\Psi$.
  
On the other hand, it is important to note that  the expression for $H$ given in Eq. (\ref{hopf-inv}) is valid {\em not only}  when the unit vector field configuration ${\bf t}(x,y,z)$ in 3D satisfies  homogeneous boundary conditions in all three  coordinates so that $R^3$ gets compactified to $S^3$, resulting in the map $S^3 \rightarrow S^2$,  {\em but also for other boundary conditions} such as periodicity in some (or all) coordinates, leading to other types of compactifications of $R^3$ \cite{jaykka}. Compact manifolds such as  $S^2 \times T^1$  and $T^3$, which lead to corresponding maps  $S^2 \times T^1 \rightarrow S^2$  and $T^3\rightarrow S^2$   are two examples. A detailed discussion of the topological aspects  of such maps may be found in \cite{jaykka}. Other examples involving compact manifolds are given in \cite{ward}.

We have studied  a  certain 3D  anisotropic, {\em inhomogeneous} Heisenberg ferromagnetic model  \cite{PLA}, where the unit vector field spin configuration  ${\bf S}(x,y,z)$ satisfies  periodic boundary conditions in the $z$ direction and homogeneous ones on the   $(x,y)$ plane. Hence it describes the map  $(S^2 \times T^1)$ $\rightarrow$ $ S^2$ \cite{jaykka}. For a special form of inhomogeneity,  we found {\em exact analytical solutions}  for   ${\bf S}(x,y,z)$ in this model. These are called
{\em hopfion vortex} solutions \cite{jaykka}.  In what follows, we use this model as an illustrative example to determine $\Psi$. We  will not repeat  all the details of this model  and its solutions since  they are described in  \cite{PLA},  but will focus  only on the parts that are relevant for determining the analytic expression for the `fixed'  twist $\Psi$  appearing in $H$.
  
  Our exact solution  for ${\bf S}(x,y,z)$  was labeled by two integers $n$ and $m$.  The three components of the emergent magnetic field 
  ${\bf B}$ for this solution were calculated using Eq. (\ref{B1}) and its cyclic permutations, by setting ${\bf t}= {\bf S}$ in them. 
  The  vector potential  ${\bf A}$  corresponding to  the above magnetic field ${\bf B}$  was  then  obtained by solving 
   $\boldsymbol {\nabla} \times{\bf A} = {\bf B}$.  Its components were found to be 
\be 
A_1 =  [\pm{1}+\cos\Theta ]\, \frac{\partial \Phi}{\partial x}; \,\, ~A_2 =  [\pm{1}+\cos\Theta ]\, \frac{\partial \Phi}{\partial y}\,\,;\, \, ~A_3 = \cos\Theta  \frac{\partial \Phi}{\partial z}.
\label{a123C}
\ee 
By substituting the respective expressions for ${\bf B}$  and ${\bf A}$   in Eq. (\ref{hopf-inv}) and integrating,  we found the  Hopf invariant  to be  an integer  given  by $H =nm$, a product of two integers. 
We also showed  that the preimages of $S^2$  for our model are knots, any two of which link exactly $H$ times. Thus the interpretation of $H$ as a linking number was also verified.

We proceed to  determine   $\Psi$   as follows. To understand how gauge fixing appears in $H$, recall that  {\bf A} is identified with {\bf V} in Eq. (\ref{AThetaPhi}), whose components are given  in  Eq. (\ref{twist_Psi}) (and its analogs).
\be
A_{i}(\Theta, \Phi)=V_{i} = \tau_{T,i} = \tau_{i} + \partial_{i} \Psi =\cos \Theta \,\,\partial_{i} \Phi + \partial_{i} {\tilde{\Psi}}.
\label{AThetaPhiTilde}
\ee  

Comparing Eq. (\ref{AThetaPhiTilde}) with  Eq. (\ref{a123C}) we had obtained for our model, we get 
\be
\frac{\partial \tilde{\Psi}}{\partial x} = \pm \frac{\partial \Phi}{\partial x};\,\,\,\,\frac{\partial \tilde{\Psi}}{\partial y} = \pm \frac{\partial \Phi}{\partial y};\,\,\,\, \frac{\partial \tilde{\Psi}}{\partial z} = 0.
 \label{Psi-x}
 \ee
In Eq. (\ref{Psi-x}), we substitute  $\tilde{\Psi}_{x} =\Psi_{x} +\beta_{x}$ as given below Eq. (\ref {twist_Psi}), where $\beta$ is defined in Eq. (\ref{beta}) (and similarly for the $y$ and $z$ derivatives of 
 $\tilde{\Psi}$). 

 We find that the  twist angle $\Psi$  is not an independent variable and  gets {\it fixed} in terms of the polar and azimuthal angles of the spin vector field 
 ${\bf S}$. This establishes   that the  twists of the space curves  which are intrinsic geometrical  quantities underlying  the  topological invariant $H$  are not arbitrary. Note that this choice of  $\Psi$ is specific to this magnetic model only.
 
 \section {{\bf  Summary of results and concluding remarks}}
In recent years, the study of topological invariants has become vital in the understanding  of  several unusual phenomena in a variety of physical systems. The well known Gauss-Bonnet relationship given in Eq. (\ref{GB1}) states that  the Euler characteristic $\chi$ (a topological invariant) is given by the  integral of the Gaussian curvature $K$ (an intrinsic geometric quantity).  The  intriguing question of whether other topological invariants such as winding numbers $W_2$ and $W_3$ and the linking number  $H$,  can  be  written down as integrals  of certain other types of intrinsic geometric quantities has not been addressed in the literature. In this paper we have shown that it is indeed possible to do so. Interestingly,  their integrands involve  torsions and twists, which are intrinsic geometric quantities for the space curves describing the manifolds concerned.  We have also found these quantities explicitly and interpreted them physically, in addition to obtaining other related results.

   Our approach which uses the {\em geometric} phenomenon of anholonomy  (or geometric phase) associated with evolving space curves is  ideal for finding the geometric quantities underlying these topological invariants. This is essentially because Frenet-Serret equations  encode the intrinsic geometry of a space curve, independent  of its position and orientation in physical space.  
   
  Our new results are as follows: (i) We have found a novel connection which shows that the topological invariants given by the winding and linking numbers are {\em identical}  to   the corresponding (appropriately normalized) global  anholonomies associated with the unit vector fields defined on the respective manifolds considered. (ii) The  integrands of these topological invariants are then shown to  emerge as    geometric quantities (which describe  intrinsic space curve geometry)  in a natural fashion. (iii) For the 2D winding number  $W_{2}$, these quantities involve the torsions of the two space curves depicting the 2D manifold. They quantify their underlying nonplanarity. (iv) In the case of the 3D winding number $W_{3}$ and the linking number $H$, we find that in addition to the three torsions, the  respective contributions from the  intrinsic twists of the evolving space curves  emerge as crucial geometric quantities.
  (v) Importantly, we find that the twist angle $\Psi$ (which plays the role of a gauge) is {\em necessary} to yield a nontrivial $W_3$ and $H$. (vi) In particular, we note that  since the target space of $W_3$ is $S^3$,   a redefined twist angle  (which we denote by $\tilde{\Psi}$) acts as a third independent angle, in addition to the angles $(\Theta,\Phi)$. On the other hand, since the target space corresponding to $H$ is $S^2$,  this angle gets {\em fixed}, leading to the twist angle $\Psi$ becoming a specific functional of $\Theta$ and $\Phi$. This functional depends on the unit vector field solution of the model concerned. (This is explained in Sec. 2.3.2.) (vii) As an illustrative example, we  have also explicitly  demonstrated how this gauge-fixing of $\Psi$ emerges in  the topological invariant $H$ for  the exact hopfion vortex soliton we have found \cite{PLA},  for the spin configuration of  a 3D anisotropic, inhomogeneous Heisenberg ferromagnet. Of the above results, those  which are significant in the context of  various types of topological solitons are summarized in Table 1 for ready reference. 
  
\begin{table}[h!]
\centering
\caption{Different mappings and corresponding topological invariants, underlying {\it curve} geometric quantities 
(torsion and twist) and the gauge for the topological solitons describing four distinct vector field textures (see text for details).}
\vskip 0.3 truecm 
\begin{tabular}{ |c|c|c|c|c|c| }
 \hline 
 {\bf Dim.} & {\bf Map} & {\bf Topological Invariant} & {\bf Geometric Quantity} & {\bf Gauge} & {\bf Topological soliton} \\
 \hline 
 2D & $S^2\rightarrow S^2$ & $W_2$ (winding no.) & Torsion & None & Magnetic skyrmion \\
 \hline 
 3D & $S^3\rightarrow S^3$ & $W_3$ (winding no.) & Torsion plus general twist & Arbitrary & Skyrmion \\ 
 \hline 
 3D & $S^3\rightarrow S^2$ & $H$ (linking no.) & Torsion plus fixed twist & Fixed & Hopfion \\ 
\hline 
 3D & $S^2\times T^1\rightarrow S^2$ & $H$ (linking no.) & Torsion plus fixed twist & Fixed & Hopfion vortex\\ 
 \hline 
\end{tabular}
\end{table} 
 
 An integral with a structure  identical to the Hopf invariant $H$ (Eq. (\ref{hopf-inv}))  has been shown to be  a topological invariant in several other fields, with  corresponding identifications of the vector field ${\bf A}={\bf V}$ as the vector potential of a  given divergence-free field  ${\bf B}$ = $\boldsymbol{\Omega}$.  In fluid dynamics, 
  $\boldsymbol{\Omega}$  is taken as a   vorticity field 
 $\boldsymbol{\omega} = \boldsymbol{\nabla}\times\boldsymbol{\bf u}$,  with its  solution being the velocity field ${\bf u}$($x,y,z$).   Moffatt \cite{moffatt1} named  the corresponding quantity  as {\em helicity}.
 Assuming the special case of localized vorticity distribution $\boldsymbol{\omega}$, which vanishes everywhere except  in two closed vortex filaments, Moffatt \cite {moffatt1}  has shown that  for this case, $H$   becomes a topological invariant  given by the {\em linking number} of  the vortex filaments in a fluid flow. In  magnetohydrodynamics, $H$ has been termed {\it magnetic helicity}, with the given local magnetic field ${\bf B}$   and its corresponding  vector potential ${\bf A}$  playing  the roles of $\boldsymbol {\Omega}$ and ${\bf V}$, respectively. Here, $H$ becomes a topological invariant  which gives the linking number of magnetic field lines.
Topological invariants resulting from similar identifications have  also appeared in  liquid crystals \cite{tai}, and in 3D static  field theories supporting solitons \cite{faddeev2,gladikowski}.  In  (2+1)D field theories \cite{wilczek}, $\boldsymbol{\Omega}$  and ${\bf V} $ are  the conserved topological current $j_{\mu}$, and its gauge potential $A_\mu$, $\mu =0,1,2$, respectively. 
 The  Hopf invariant has also been studied in Hamiltonian systems \cite{arnold}, and its relationship with helicity in fluid dynamics has been discussed.  Thus, our analysis should be useful in identifying the   geometric quantities appearing in the integrands of  topological invariants appearing in diverse  physical  systems, especially those systems in which space curves appear as  basic entities. Vortex filaments in fluid dynamics, curved field lines in magnetic systems and liquid crystals, phase space trajectories (in 3D) of a Hamiltonian system, elastic filaments, biopolymers, etc., are some examples of space curves.
 
As already mentioned in  Sec.~1.3, there is a direct correspondence between the concept of parallel transport of {\em classical vectors} and that of {\em quantum states}  \cite{shapere}.
 Considering the vectors ${\bf N}$ and ${\bf B}$ given in Eq. (\ref{NB}), we define a  normalized  classical  {\em complex} vector  ${\bf M}=  \frac{1}{\sqrt{2}} ({\bf N} + i {\bf B}) \exp( -i \int \tau_{T,1} (x)\, dx)$ and map it  to a normalized quantum state 
 $|\psi\rangle$ in  Hilbert space. Using this expression for ${\bf M}$, we can show that   the {\em classical} Berry connection $\tau_{T,1}(x)$, which  is the {\em total twist density} of a space curve, [see  Eqs. (\ref{twist}) and (\ref{V1})] takes  on the  form $i{\bf M}^{*}\cdot d{\bf M}/dx$, where  ${\bf M}^{*}$ is the complex conjugate of ${\bf M}$. This  has the same form as the quantum Berry connection $ i\langle\psi |\frac{\partial}{\partial  x} \psi\rangle$ \cite{berry,aharanov}, and may be regarded as its classical analog.
 
 The deep connection between topology and the quantum Hall effect conductance is by now well established. In models like 2D Chern insulators, a unit vector field ${\bf t} (k_x,k_y)$  can be constructed in the 2D Brillouin zone (momentum space)  in terms of the  Hamiltonian  of the model \cite{chern}. 
 Here, the target space is a Bloch sphere $S^2$ while the physical space is compactified to the torus $T^2$ due to periodic boundary conditions of the quantum wave function in the 2D momentum space. The map $T^2 \rightarrow S^2$  is classified by an integer topological invariant, which is called {\em Chern number} \cite{chern1} in the quantum context.
 
 As explained in Sec. (2.2),   the map $S^2 \rightarrow S^2$ is classified by the winding number $W_2$  given in Eq. (\ref{W2}) in terms of angle variables. By using the representation for ${\bf t}$  given in Eq. (\ref{t}), a short calculation shows  that  Eq. (\ref{W2}) can also be written in the form
 \be
 W_2= \frac{1}{4\pi} \int\int\, \mathbf{t} \cdot
 (\partial_{y}\mathbf{t} \times \partial_{z}\mathbf{t})\,\,dy\,\, dz.
 \label{W2t}
 \ee
 It is interesting to note that   the integral expression of  the Chern number \cite{chern1}  is  {\em identical}  to that of  the winding number $W_{2}$  given in Eq. (\ref{W2t}), when  the partial derivatives of ${\bf t}$ with respect to physical space variables $(x,y)$ in the integrand are replaced by the corresponding derivatives with respect to the periodic momentum space variables $(k_{x}, k_{y})$, and the integration is over the 2D Brillouin zone. Furthermore, as is well known, Chern number is the integral over (quantum) Berry curvature. As
stated above Eq.~(\ref{O1angles}) and below Eq.~(\ref{W2torsion}), we have found that $W_2$  is an integral over classical Berry curvature, by using the anholonomy of space curves. Hence the  winding number  $W_2$ may be regarded as the  classical analog of the Chern number.
 
  More recently, topological insulators in 3D, called Hopf insulators \cite{moore} have  attracted a lot of attention. Their Hamiltonians can be associated with a map $T^3 \rightarrow S^2$. Such a map is classified by an integer, which is just the Hopf invariant arising  from the periodicity  in the 3D Brillouin zone momentum space in the quantum context. Topological invariants are known to play a crucial role in understanding various  physical phenomena in condensed matter systems. Hence it would be interesting to understand and physically interpret  the quantum analogs of the  classical  geometric quantities describing space curves that we have identified as the integrands of topological invariants such as winding numbers and the linking number, by considering  specific physical models.
 
 We thank Saikat Banerjee for help with the figure. The work  of A. S. at Los Alamos National Laboratory was carried out under the auspices of  the U.S. DOE and NNSA under Contract No. DEAC52-06NA25396.

\end{document}